\documentclass[aps,prb,twocolumn,notitlepage,showpacs,floatfix,superscriptaddress]{revtex4-1}
\usepackage{dcolumn}
\usepackage{tabularx}
\usepackage{bm}
\usepackage{soul}
\usepackage{amsmath,amssymb,graphicx}
\usepackage[colorlinks=true,citecolor=blue,urlcolor=blue,linkcolor=blue]{hyperref}
\usepackage{environ}

\NewEnviron{eqnsplit}{%
\begin{equation}
\begin{split}
  \BODY
\end{split}
\end{equation}
}

\newcommand{\mrm}[1]{\mathrm{#1}}

\newcommand{\bmk}{{\bm k}}
\newcommand{\bmr}{{\bm r}}

\begin{document}
\title{Coherent Excitonic Quantum Beats in Time-Resolved Photoemission Measurements}

\author{Avinash Rustagi}
\email{arustag@purdue.edu}
\affiliation{Department of Physics, North Carolina State University, Raleigh, NC 27695}
\affiliation{School of Electrical and Computer Engineering, Purdue University, West Lafayette, IN 47907}
\author{Alexander F. Kemper}
\email{akemper@ncsu.edu}
\affiliation{Department of Physics, North Carolina State University, Raleigh, NC 27695}
\date{\today}
\begin{abstract}
Coherent excitation of materials via ultrafast laser pulses can have interesting, observable dynamics in time-resolved photoemission measurements. The broad spectral width of ultrafast pump pulses can coherently excite multiple exciton energy levels. When such coherently excited states are probed by means of photoemission spectroscopy, interference between the polarization of different exciton levels can lead to observable coherent exciton beats. Here, we present the theoretical formalism for evaluating the Time- and Angle- Resolved Photoemission Spectra (tr-ARPES) arising from the coherently excited exciton states. We subsequently apply our formalism to a simple model example of hydrogenic exciton energy levels to identify the dependencies that control the quantum beats. Our findings indicate that the most pronounced effect of coherent quantum excitonic beats is seen midway between the excited exciton energy levels and the central energy of the pump pulse provides tunability of this effect.
\end{abstract}

\maketitle
\section{Introduction}
Light-matter interaction is fundamental to probe the properties of materials both in and out of equilibrium. Light induced polarization in matter leads to interesting and relevant coherent and incoherent phenomena. Studies based on this interaction has revealed a plethora of information in regards to the excited quantum states\cite{luSC_Arpes,chenTI_Arpes}, the coupling between various degrees of freedom \cite{Gerber71}, and the dynamical time scales associated with fundamental processes in materials\cite{Arpes_review2016}. While this interaction is significant to explaining fundamental coherent phenomena like entanglement\cite{Lenihan2002}, inversion-less lasing\cite{Scully1989}, and Rabi oscillations of excitons \cite{ShamPRL_2001}, it also has applications in optoelectronics\cite{PlankenTHz1992}.\\

Coherent quantum beats is an important spectroscopic signature, providing information about excited quantum states. These are coherent in the sense of simultaneously excitating two or more discrete excited energy levels, thereby creating a superposition quantum state, which in turn leads to interference between the time-dependent polarizations of these excited levels. These quantum beats are observed as periodic oscillations (with period given by the inverse transition energy between the excited levels) in time-domain measurements and are particularly useful in the understanding of coherent light-matter interaction. Quantum beats have been observed for different time resolutions through control of the excitation pump pulse duration and has been particularly important in measuring molecular constants. Shorter temporal pulses with wider frequency spectrum have been useful in monitoring the real time vibrational dynamics\cite{VibDyan_PRL_2002}, and the transition states in molecules\cite{QB_TransitionState}. On the other hand, longer temporal pulses with narrower frequency spectrum has applications in determining molecular structural parameters like spin orbit coupling constant\cite{QB_SO_1985}, and dipole moments in excited states\cite{QB_DipMom_1988,Huber_MolQB_1992}. Typically, coherent excitonic quantum beats have been observed in diffraction/absorption/transmission based optical measurements like Four Wave Mixing (FWM) and photon echo experiments on semiconductor quantum wells following photoexcitation by a ultrafast laser pulse\citep{LeoPRL_1991,koch1992beats,haacke1997resonant,PRB2015}, in hybrid organic-inorganic perovskites \cite{SpinExBeat2017}, and most recently in atomically thin layer of RSe$_2$\citep{Beats_RSe2_2018}.  \\

One of the disadvantages of the diffraction/absorption/transmission based optical measurements is that the observations made are momentum averaged. Time- and Angle- Resolved Photoemission  Spectroscopy (tr-ARPES) is known to provide energy, momentum, and time resolution\citep{Damascelli_RMP2003}. This novel experimental technique thus provides complimentary information to the conventional optical methods. Femtosecond ultrafast laser pulses can simultaneously excite multiple exciton states into a coherent state. The photoemission intensity from the coherent state can show exciton beats with frequency set by the difference in exciton energies. Transient exciton creation and their subsequent dynamics have been observed in both metals and semiconductors \cite{petek2014transient, silkin2015ultrafast}. Analogous coherent beat oscillations in time-resolved Two-photon photoemission (2PPE) on metal surface has been observed \citep{hofer1997time}, where the photoelectron couples to its image charge partner forming a bound state. \\

Recent studies on the signatures of both coherent and incoherent excitons via tr-ARPES  at varying levels of complexity have raised interesting possibilities \citep{perfetto2016first,rustagi2018ARPES,Sangalli2018,Nasu_2017Exciton}. The potential of resolving excitons through tr-ARPES opens up interesting opportunities of characterizing them by means of the momentum resolved photoemission spectrum. The coherent excitation of the exciton states has raised the possibility of observing the coherent excitonic quantum beats via photoelectron spectroscopy and is the focus of this paper. \\

Here, we present our work evaluating the signatures of coherent exciton beats observed in tr-ARPES. This paper is organized as follows: In Sec.~\ref{Coherent Exciton State}, we set up the formalism for coherent exciton state generation by the pump pulse. Following which, Sec.~\ref{Photoemission Theory} applies the semi-perturbative theory of photoemission to the coherent exciton state. In Sec.\ref{Application to a simple example}, we apply the developed theory to a model example calculation where the lowest two Rydberg exciton states are excited by the pump and subsequently probed by photoemission spectroscopy. Finally, we summarize our conclusions in Sec.~\ref{Conclusions}.

\section{Theoretical Formalism}
\label{Theoretical Formalism}
%
%
\begin{figure}[htbp]
\centering
\includegraphics[width=0.45\textwidth]{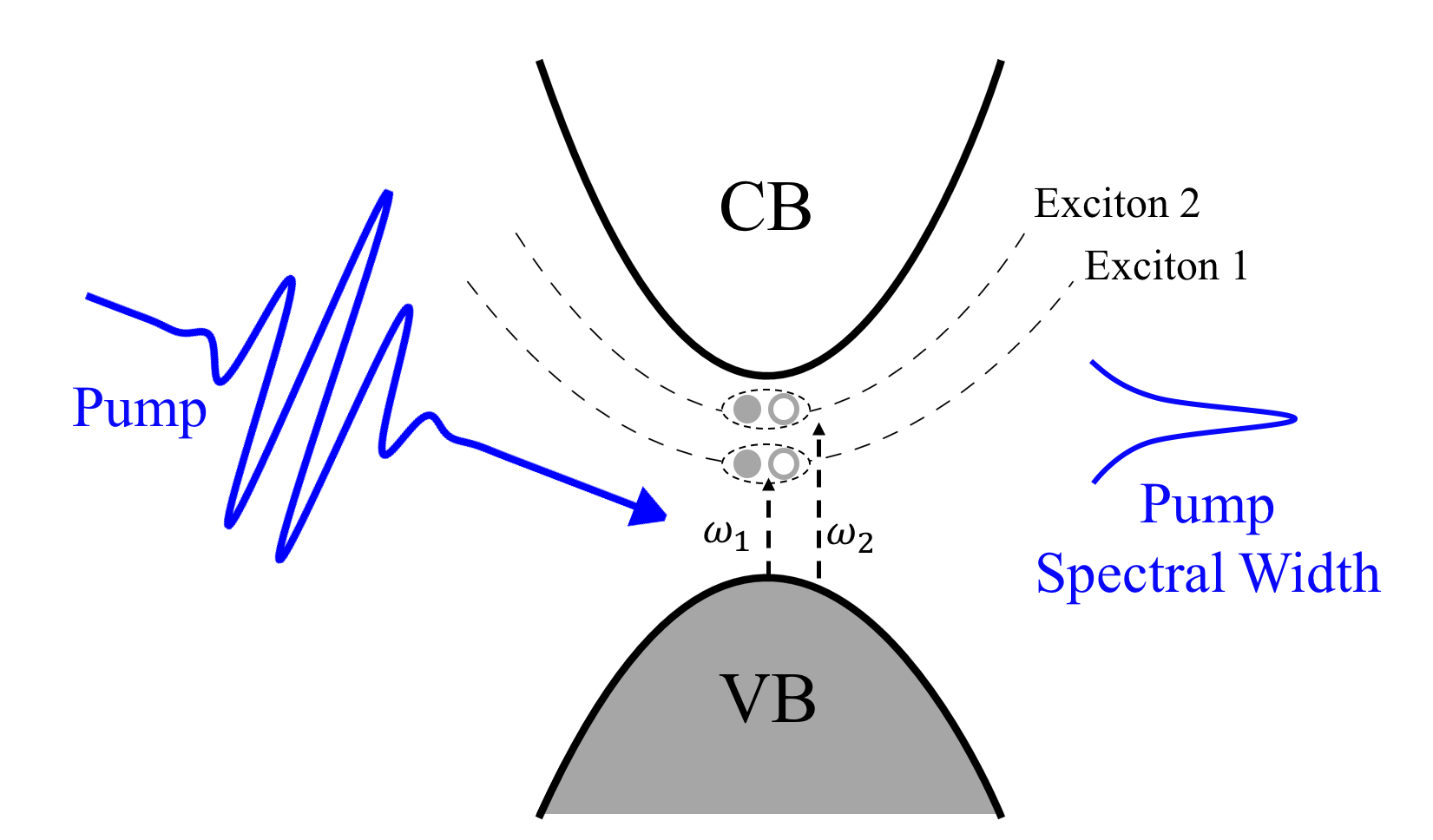}
\caption{\label{schematic}Schematic of the excited exciton levels by the ultrashort pump pulse (large spectral width).}
\end{figure}
%
%
Excitons are bound states of electron-hole pairs, which dominate the sub-band gap optical spectrum of a semiconductor in addition to the unbound electron-hole pairs at energies above the band gap. These bound composite particles along with the unbound electron-hole pairs form the excitation spectrum of the semiconductor. The composite Bosonic excitations can be directly excited by the pump pulse for appropriate choice of laser energy (sub-band gap). In such a scenario, we can effectively describe the system by an effective system of excitons coupling to the optical pump field. Such an effective Bosonic model describing a system of interacting Fermions works well in the limit of low density and temperature \cite{RochatPRB_2000,TassonePRB_2001}.
\subsection{Coherent Exciton State}
\label{Coherent Exciton State}
With the advent of ultrashort femtosecond pulses which have a broad energy spectrum, it is possible to coherently excite multiple exciton levels simultaneously (see Fig.~\ref{schematic}). For simplicity we can assume that there are two exciton states $\vert 1 \rangle$ and $\vert 2 \rangle$ with energies $\omega_1$ and $\omega_2$. The Hamiltonian describing the sub-band gap composite Bosonic excitons coupling to the electromagnetic (EM) field is \citep{Hanamura_ExH1977,Jorda_ExPhInt1993,Ostreich1998,fernandez1999exciton}
\begin{equation}
H= H_0 - V E(t) P 
\end{equation}
where the unperturbed Hamiltonian describes the exciton energy levels and the unbound electrons and holes in the Conduction and Valence bands ($H_\mrm{eh}$)
\begin{equation}
H_0 =  \omega_1 A_1^\dagger A_1 +  \omega_2 A_2^\dagger A_2 + H_\mrm{eh}
\end{equation}
The polarization $P$ that couples to the applied EM field considering that the optical field energy is sub-band gap, is 
\begin{equation}
P=\dfrac{G_1}{\sqrt{V}} \left( A_1^\dagger + A_1 \right) + \dfrac{G_2}{\sqrt{V}} \left( A_2^\dagger + A_2 \right)
\end{equation}
where $V$ is the volume of the system.
%
%
\begin{figure}[htbp]
\centering
\includegraphics[width=0.45\textwidth]{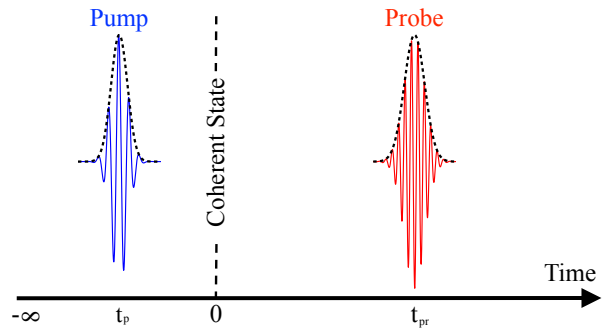}
\caption{\label{timeline}Timeline of the pump-probe photoemission measurement. The pump acts at earlier times creating a coherent state at time $t=0$, following which the probe pulse is used for measurements.}
\end{figure}
%
%
Here $A_i^\dagger/A_i$ correspond to the creation/annihilation of the composite Bosonic exciton $i=\{1,2\}$, and $G_1/G_2$ is the electric dipole matrix element corresponding to coupling of photons to excitons labeled `$1/2$'. Since we are considering coherent optical excitation of excitons by photons, the excited excitons are predominantly the ones with zero center of mass momenta. Thus the composite exciton operators can be written in terms of the fundamental electron-hole operators\cite{Elliott1957}
\begin{equation}
A_i^\dagger = \sum_{\bm p} \phi_{i {\bm p}} b^\dagger_{\bm p} a_{\bm p}
\end{equation}
where $b^\dagger_{\bm p}$ corresponds to the creation of electron in conduction band (CB), $a_{\bm p}$ corresponds to the annihilation of electron in valence band (VB) i.e. creation of a hole, and $\phi_{i {\bm p}}$ is the envelope wavefunction for the exciton eigenstate. The solution to the time-dependent Schrodinger equation with Hamiltonian $H$ when the pump excites only exciton states is expressed as\citep{Glauber1963}
\begin{equation}
\begin{split}
\vert \Psi(t) \rangle &= e^{-i H_0 t } \prod_{i=\{1,2\}} e^{ i K_i(t) A_i^\dagger} e^{ -\vert K_i(t) \vert^2 /2 } \vert 0 \rangle \\
\end{split}
\end{equation}
where $ K_{i}(t) =  \sqrt{V} G_i  \int_{-\infty}^{t} dt' \,  E(t') e^{i\omega_i t'}$.\\

The pump pulse of temporal width $\sigma_{p}$ and central frequency $\Omega$, centered around time $t_p$ is
\begin{equation}
E(t')=E_0 e^{-(t'-t_p)^2/2\sigma_p^2} \cos[\Omega(t'-t_p)]
\end{equation}

Following the timeline shown in Fig.~\ref{timeline}, the pump acts at some earlier time $t_p$ (such that there is no overlap with the probe pulse) and forms the coherent state. Since the pump is narrow, $K_i$ are constant for all positive times
\begin{equation}
\label{coherenteigval}
\begin{split}
K_{i}(t=0) &=  \sqrt{V} G_i  \int_{-\infty}^{0} dt' \,  E(t') e^{i\omega_i t'} \\
&\approx  \sqrt{V} G_i  \int_{-\infty}^{\infty} dt' \,  E(t') e^{i\omega_i t'} \\
&\approx  \sqrt{\dfrac{\pi}{2}} E_0 \sqrt{V} G_i  \sigma_p  e^{i\omega_i t_p} e^{-\sigma_p^2 (\omega_i -\Omega)^2/2} \\
\end{split}
\end{equation}
where we have neglected the negligibly small exponential term $\sim e^{-\sigma_p^2 (\omega_i + \Omega)^2/2}$. Since we consider the coherent signatures of exciton in time-resolved photoemission measurements, we do not consider any loss of coherence. \\

The coherent state formed by the pump is our starting point for the photoemission spectrum evaluation. Thus, the wavefunction of the exciton coherent state at any finite time $t>0$ is
\begin{equation}
\begin{split}
\vert \Psi(t) \rangle &=N e^{-i H_0 t } e^{ i K_1A_1^\dagger}  e^{ i K_2A_2^\dagger} \vert 0 \rangle
\end{split}
\end{equation}
where the normalization $N= e^{ -\vert K_1 \vert^2 /2 } e^{ -\vert K_2 \vert^2 /2 } $. The wavefunction can be further simplified by expressing the coherent state in number-basis 
\begin{equation}
\begin{split}
\vert \Psi(t) \rangle &=N e^{-i H_0 t } e^{ i K_1A_1^\dagger}  e^{ i K_2A_2^\dagger} \vert 0 \rangle \\
&=N  \sum_{n_1,n_2} \dfrac{(i \bar{K}_1(t))^{n_1}}{\sqrt{n_1 !}} \dfrac{(i \bar{K}_2(t))^{n_2}}{\sqrt{n_2 !}} \vert n_1;n_2 \rangle \\
&= N e^{i \bar{K}_1(t) A_1^\dagger} e^{i \bar{K}_2(t) A_2^\dagger} \vert 0 \rangle
\end{split}
\end{equation}
where $\bar{K}_{1/2}(t) = K_{1/2} e^{-i\omega_{1/2}t}$. 

\subsection{Photoemission Theory}
\label{Photoemission Theory}
The general theory of photoemission \cite{freericks_ARPES} which was previously applied to study the contribution of incoherent excitons in photoemission measurements \citep{rustagi2018ARPES} is now used to evaluate the tr-ARPES spectra for coherent excitons. 
\begin{equation}
\vert \Psi (t) \rangle = U(t,t_0) \vert \Psi(t_0)\rangle
\end{equation}
is the time dependent wavefunction due to the effect of the pump governed by the time-evolution operator
\begin{equation}
U(t,t_0) = \mathcal{T}_t \exp \left( -\dfrac{i}{\hbar} \int_{t_0}^{t} dt' H_{pump}(t')\right)
\end{equation}
In presence of both pump and probe, the wavefunction is given by
\begin{equation}
\vert \Psi_F (t) \rangle = \bar{U}(t,t_0) \vert \Psi(t_0)\rangle
\end{equation}
where the time-evolution operator for pump + probe fields
\begin{equation}
\bar{U}(t,t_0) = \mathcal{T}_t \exp \left( -\dfrac{i}{\hbar} \int_{t_0}^{t} dt' [H_{pump}(t')+H_{probe}(t')]\right)
\end{equation}
Assuming the probe pulse to be weak and linearizing the time-evolution operator
\begin{equation}
\bar{U}(t,t_0) \approx U(t,t_0) - \dfrac{i}{\hbar} \int_{t_0}^{t} dt' \, U(t,t') H_{probe}(t') U(t',t_0)
\end{equation}
where for photoemission, the probe Hamiltonian annihilates a CB electron ($b_{\bm k'}$) and creates a free photoelectron ($f^\dagger_{\bm k}$)
\begin{equation}
H_{probe} (t)= s(t) e^{-i\omega_0 t} M_{{\bm k},{\bm k'}} f^\dagger_{\bm k} b_{\bm k'} 
\end{equation}
where ${\bm k}=\{k_{||},k_z\}$ and ${\bm k'}=\{k_{||},k'_z\}$, thereby conserving the parallel component of momenta but not the perpendicular one.  Given the uncertainty in the z-component of momentum, there is an unknown offset in the perpendicular momentum. Therefore, we set $k'_z=0$ noting that with variations of the probe photon energy, the c-axis dispersion can be mapped given the unknown offset. The probe pulse is centered around energy $\omega_0$ and has a temporal profile $s(t)$. To probe the time evolution of the non-equilibrium system, the probe pulse is applied at different delay times and the photoemission intensity is measured. The probability to find the photoemitted electron with momentum $k$ in a solid angle $d\Omega_k$
\begin{equation}
I(t)= \lim_{t\rightarrow \infty} \dfrac{k^2dkd\Omega_k}{(2\pi)^3} P(t) ; \,\, P(t) = \sum_{m} \vert \langle \Psi_m^{1<}; {\bm k} \vert \Psi_F (t) \rangle\vert^2
\end{equation}
Therefore
\begin{equation}
\begin{split}
P(t)&=\sum_{m} \vert \langle \Psi_m^{1<}; {\bm k} \vert \Psi_F(t) \rangle\vert^2 \\
&=\sum_{m}  \langle \Psi_F(t) \vert \Psi_m^{1<}; {\bm k} \rangle \langle \Psi_m^{1<}; {\bm k} \vert \Psi_F(t) \rangle \\
&=  \int_{t_0}^{t} dt_1 \,  \int_{t_0}^{t} dt_2\, s(t_1) s(t_2) e^{i\omega_0 (t_2 -t_1)} e^{-i (\omega_e+W)(t_2 -t_1)} \\
&\times \vert M_{{\bm k},{\bm k'}}\vert^2 \langle \Psi(t_2)\vert  b^\dagger_{\bm k'} U(t_2,t_1) b_{\bm k'}  \vert \Psi(t_1)\rangle
\end{split}
\end{equation}
where $\omega_e$ is the kinetic energy of the photoemitted electron while $W$ is the energy lost in overcoming the work function of the material. We consider the photoemission process to be instantaneous which is a valid assumption for high energy photons i.e. the \textit{sudden} approximation \citep{Damascelli_RMP2003}. The state $\vert \Psi_m^{1<}; {\bm k} \rangle = \vert \Psi_m^{1<} \rangle \otimes \vert {\bm k} \rangle $ is the direct product of the material wavefunction with one less electron $\vert \Psi_m^{1<} \rangle$ and the free photoemitted electron with momenta ${\bm k}$.\\

It is important to highlight that in the following evaluation of the photoemission spectra, the coherent state eigenvalue $K_i$ is taken to be constant since we have considered that the pump acts at earlier times. This makes it easier to calculate the spectra, however we should note that the results of the subsequent calculations with time-independent eigenvalue $K_i$ are valid only when the probe pulse acts after the duration of the pump pulse and there is no overlap between the two. The expression for the photoemission intensity involves the probe temporal profile which is centered about $t_{pr}>0$, thus we can simply move forward with the calculation assuming $t_1,t_2 >0$. We now consider the evaluation of the ARPES spectra by evaluating the action of the CB electron annihilation from the coherent exciton state, given by \\
\begin{equation}
\begin{split}
b_{\bm k'} \vert \Psi(t_1) \rangle &= i N \left[ \bar{K}_2(t_1) \phi_{2 \bmk'} + \bar{K}_1(t_1) \phi_{1 \bmk'} \right] \\
& \quad \times a_{\bmk'}e^{i \bar{K}_1(t_1) A_1^\dagger} e^{i \bar{K}_2(t_1) A_2^\dagger} \vert 0 \rangle   \\
\end{split}
\end{equation}
using the commutation relation
\begin{equation}
\begin{split}
&[b_{\bm k'}, e^{i \bar{K}_i(t) A_i^\dagger} ] = i \bar{K}_i(t) \phi_{i \bmk'} e^{i \bar{K}_i(t) A_i^\dagger} a_{\bmk'} \quad i=\{1,2\}. \\
\end{split}
\end{equation}

Now we consider the next step in evaluation of the required matrix element to get the action of time-evolution operator $U(t_2,t_1)$ on the CB electron annihilated coherent exciton state $b_{\bm k'} \vert \Psi(t_1) \rangle$. Here we note that the choice of no overlap between the pump and probe pulses and the temporal profile dependence of the photoemission intensity implies that $U(t_2,t_1) = e^{-iH_0 (t_2 - t_1)}$ since $t_1, t_2 >0$. Therefore
\begin{equation}
\begin{split}
&U(t_2,t_1) b_{\bm k'} \vert \Psi(t_1) \rangle  = i N \left[ \bar{K}_2(t_1) \phi_{2 \bmk'} + \bar{K}_1(t_1) \phi_{1 \bmk'} \right] \\
&\times e^{i \varepsilon_{v,\bmk'} (t_2-t_1) }  e^{i \bar{K}_1(t_2) A_1^\dagger} e^{i \bar{K}_2(t_2) A_2^\dagger}  a_{\bmk'} \vert 0 \rangle  \\
\end{split}
\end{equation}
where the state $a_{\bmk'} \vert n_1;n_2 \rangle$ has $n_1/n_2$ excitons in level $1/2$ and an absence of electron with momentum ${\bm k'}$ from the VB. Thus, we set the unperturbed energy of this state to be $n_1 \omega_1 + n_2 \omega_2 - \varepsilon_{v,\bmk'}$ where $\varepsilon_{v,\bmk'}$ is the energy of the VB state with the absent electron. Hence the matrix element is
\begin{widetext}
\begin{equation}
\begin{split}
\langle  \Psi(t_2) \vert b_{\bm k'}^\dagger U(t_2,t_1) b_{\bm k'} \vert \Psi(t_1) \rangle  &= N^2 \left[ \bar{K}^*_2(t_2) \phi^*_{2 \bmk'} + \bar{K}^*_1(t_2) \phi^*_{1 \bmk'} \right]  \left[ \bar{K}_2(t_1) \phi_{2 \bmk'} + \bar{K}_1(t_1) \phi_{1 \bmk'} \right] e^{i \varepsilon_{v,\bmk'} (t_2-t_1) } \\
& \times \langle 0 \vert a_{\bmk'}^\dagger e^{-i \bar{K}^*_2(t_2) A_2} e^{-i \bar{K}^*_1(t_2) A_1} e^{i \bar{K}_1(t_2) A_1^\dagger} e^{i \bar{K}_2(t_2) A_2^\dagger}  a_{\bmk'} \vert 0 \rangle 
\end{split}
\end{equation}
\end{widetext}
%
%
\begin{figure*}[hbtp]
\centering
\includegraphics[scale=0.5]{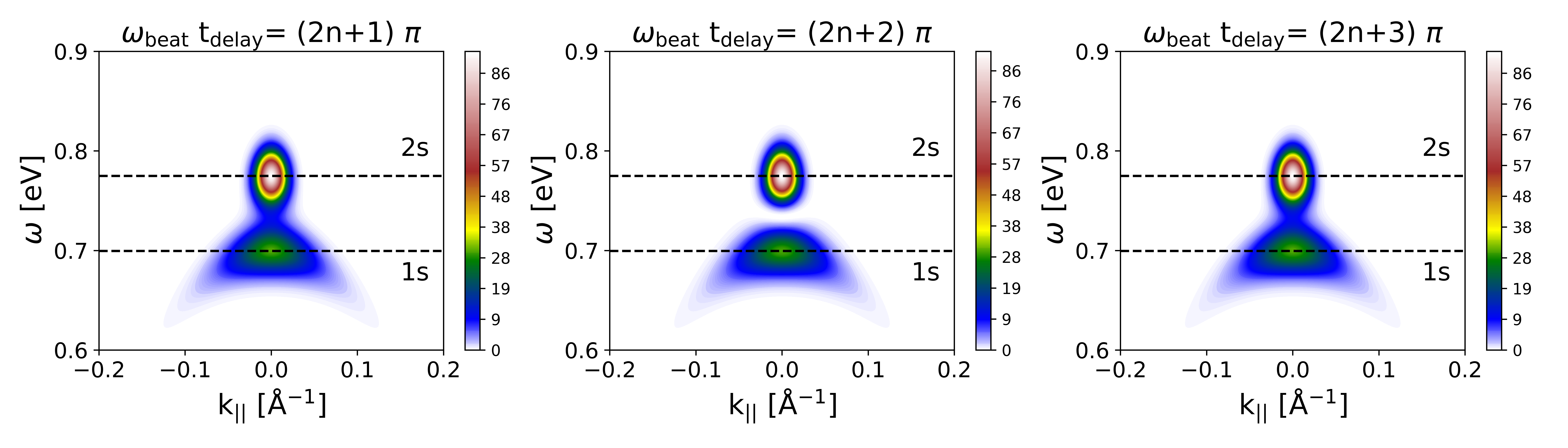}
\caption{\label{ARPES}ARPES Spectra at different delay times. The different time snapshots display the time dynamics of the coherent exciton state. The region where the beat oscillations are most pronounced is at the energies between the two exciton energies marked by dashed horizontal lines.}
\end{figure*}
%
%
Using the commutation relation
\begin{equation}
\begin{split}
[e^{-i \bar{K}^*_i(t_2) A_i},  a_{\bmk'}] &= i \bar{K}^*_i(t_2) \phi_{i \bmk'}^* e^{-i \bar{K}^*_i(t_2) A_i}  b_{\bmk'}\\
\end{split}
\end{equation}
and the Baker-Campbell-Hausdorff formula 
\begin{equation}
\begin{split}
&e^{-i \bar{K}^*_i(t_2) A_i} e^{i \bar{K}_i(t_2) A_i^\dagger} = e^{i \bar{K}_i(t_2) A_i^\dagger} e^{-i \bar{K}^*_i(t_2) A_i} e^{\vert \bar{K}_i(t_2)\vert^2} \\
\end{split}
\end{equation}
we can find the matrix element in photoemission intensity to be
\begin{equation}
\begin{split}
\langle & \Psi(t_2) \vert b_{\bm k'}^\dagger U(t_2,t_1) b_{\bm k'} \vert \Psi(t_1) \rangle  =  e^{i \varepsilon_{v,\bmk'} (t_2-t_1) }  \\
&\times \left[ \bar{K}_2(t_1) \phi_{2 \bmk'} + \bar{K}_1(t_1) \phi_{1 \bmk'} \right]  \left[ \bar{K}^*_2(t_2) \phi^*_{2 \bmk'} + \bar{K}^*_1(t_2) \phi^*_{1 \bmk'} \right] \\
\end{split}
\end{equation}
which clearly indicates four contributing terms to the photoemission intensity. The two terms of the form $\bar{K}^*_2(t_2)  \bar{K}_2(t_1) \vert \phi_{2 \bmk'} \vert^2$ and $\bar{K}^*_1(t_2) \bar{K}_1(t_1) \vert \phi_{1 \bmk'}\vert^2$ capture the individual contributions from the exciton levels where the mixing terms of the form $\bar{K}^*_2(t_2)  \bar{K}_1(t_1) \phi^*_{2 \bmk'} \phi_{1 \bmk'} $ and $\bar{K}^*_1(t_2)  \bar{K}_2(t_1) \phi^*_{1 \bmk'} \phi_{2 \bmk'} $ capture the interference between the polarizations of the two excited exciton levels.\\

Assuming the probe temporal profile to be Gaussian center around time $t_{pr}$ with temporal width $\sigma$,
\begin{equation}
s(t) = e^{-\dfrac{(t-t_{pr})^2}{2 \sigma^2}}
\end{equation}
We can take the long time limit for the integrals $t=\infty$ and the initial time $t_0=-\infty$ since the probe pulse is narrow and the integral measure will be predominantly zero excluding the region $t_{pr} - 5\sigma<t_{1/2}<t_{pr} + 5\sigma$. We can evaluate the time integrals by Wigner transforming the time arguments into average time $t_a = (t_2 + t_1)/2 $ and relative time $ t_r =t_2 -t_1$.\\

The probe pulse profile can be expressed as
\begin{equation}
\begin{split}
s(t_1) s(t_2) &= e^{-\dfrac{(t_a-t_{pr})^2}{ \sigma^2}} e^{-\dfrac{(t_r)^2}{4 \sigma^2}}
\end{split}
\end{equation}
%
%
\begin{figure*}[hbtp]
\centering
\includegraphics[scale=0.5]{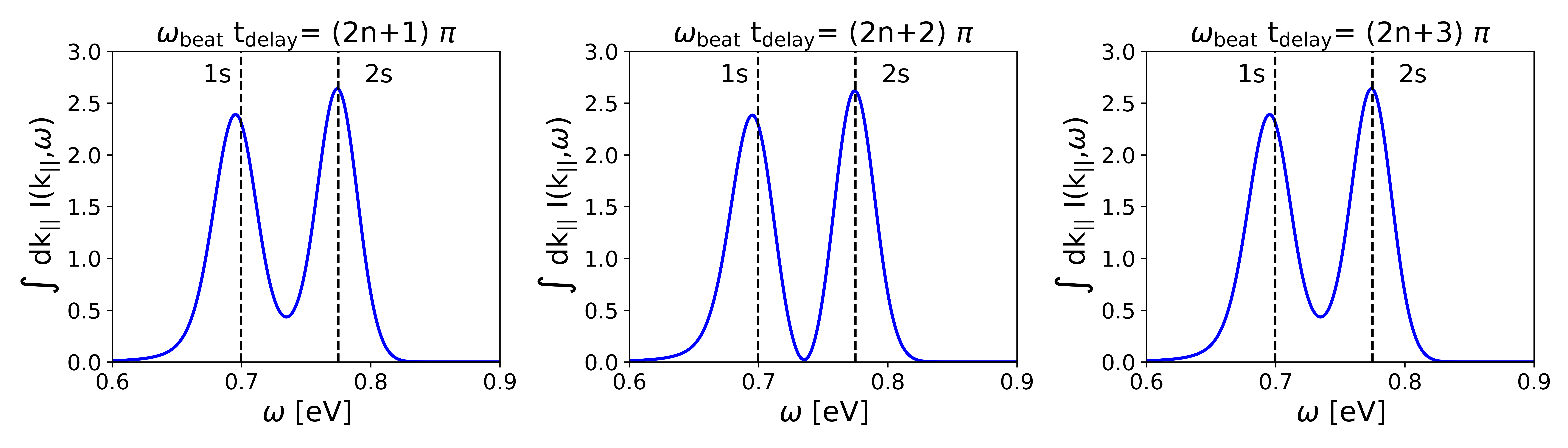}
\caption{\label{kint_ARPES}Momentum-integrated spectra at different delay times displaying the oscillation in the ARPES intensity which is most prominent in the energy range between the two exciton energies.}
\end{figure*}
%
%

Hence
\begin{widetext}
\begin{equation}
\label{offdiag_PES}
\begin{split}
P (t_d) &= \int_{-\infty}^{\infty} dt_1 \,  \int_{-\infty}^{\infty} dt_2\, s(t_1) s(t_2) e^{i\omega_0 (t_2 -t_1)} e^{-i (\omega_e+W)(t_2 -t_1)} \vert M_{{\bm k},{\bm k'}}\vert^2  \langle  \Psi(t_2) \vert b_{\bm k'}^\dagger U(t_2,t_1) b_{\bm k'} \vert \Psi(t_1) \rangle \equiv \sum_{i,j={1,2}} P_{ij} \\
\end{split}
\end{equation}
\end{widetext}
where $P_{ij}$ is the contribution to photoemission spectra from the individual exciton levels ($i=j$) and the cross-terms i.e. interference terms ($i \neq j$). These are given by
\begin{widetext}
\begin{equation}
\label{offdiag_PESeval}
\begin{split}
P_{ij} &= \int_{-\infty}^{\infty} dt_1 \,  \int_{-\infty}^{\infty} dt_2\, s(t_1) s(t_2) e^{i\omega_0 (t_2 -t_1)} e^{-i (\omega_e+W)(t_2 -t_1)} \vert M_{{\bm k},{\bm k'}}\vert^2 \bar{K}^*_i(t_2)  \bar{K}_j(t_1) \phi^*_{i \bmk'} \phi_{j \bmk'} \\
&= 2\pi \sigma  \vert M_{{\bm k},{\bm k'}} \vert^2  K_i^* K_j \phi_{i \bmk'}^* \phi_{j \bmk'}  e^{i (\omega_i -\omega_j) t_{pr}} e^{-\sigma^2 (\omega_i -\omega_j)^2 / 4}  e^{-\sigma^2 (\omega_0 - \omega_e - W + \varepsilon_{v,{\bm k'}} +  (\omega_i+\omega_j)/2)^2}
\end{split}
\end{equation}
\end{widetext}
The interference between the polarizations of the two exciton levels leads to cross terms $P_{12}$ and $P_{21}$. The significance of the interference comes in the harmonic time dependence with frequency set by the energy difference of the exciton levels. There are two linewidth like decaying exponential factors suggesting two physical implications. The exponential with the difference in the exciton energies incorporates the fact that the interference effect probed is suppressed when the exciton levels are far apart in energy. The other exponential that looks like energy conservation suggests that the most pronounced effect of the interference term is midway between the exciton levels.

The incoherent exciton contributions to the photoemission spectra are accounted for by the terms $P_{11}$ and $P_{22}$. These were discussed in previous theoretical studies on the exciton contribution to ARPES in semiconducting materials \citep{rustagi2018ARPES,Nasu_2017Exciton}. The key aspects of the contributions is that they have spectral intensity below the conduction band displaced by the exciton binding energy and their width in momentum is controlled by their corresponding wavefunction spread.

The cross term contributions are from the interference of polarizations between the two exciton levels. Since the two cross terms involve different exciton levels $i \neq j$, there is a harmonic time dependence as seen in Eq. \ref{offdiag_PESeval}.

The coherent state eigenvalues $K_{i=\{1,2\}} $ (Eq.\ref{coherenteigval}) can be expressed as
\begin{equation}
\begin{split}
K_{i=\{1,2\}} \equiv \kappa_{i=\{1,2\}} e^{i\omega_i t_p}
\end{split}
\end{equation}
and assuming that the wavefunctions are real, then the photoemission spectra is
\begin{widetext}
\begin{equation}
\label{FullARPES}
\begin{split}
P (t_d) &= \sum_{i={1,2}} 2\pi \sigma  \vert M_{{\bm k},{\bm k'}}   \vert \kappa_i^2  \phi_{i \bmk'}^2  e^{-\sigma^2 (\omega_0 - \omega_e - W + \varepsilon_{v,{\bm k'}} +  \omega_i)^2} \\
& +  4 \pi \sigma  \vert M_{{\bm k},{\bm k'}} \vert^2  \kappa_1 \kappa_2 \phi_{1 \bmk'} \phi_{2 \bmk'}  \cos \left[(\omega_2-\omega_1) (t_{pr}-t_p)\right] e^{-\sigma^2 (\omega_1 -\omega_2)^2 / 4}  e^{-\sigma^2 (\omega_0 - \omega_e - W + \varepsilon_{v,{\bm k'}} +  (\omega_1+\omega_2)/2)^2}
\end{split}
\end{equation}
\end{widetext}
where the oscillatory term depends on the difference between the probe and the pump $t_{pr}-t_p \equiv t_\mathrm{delay}$ i.e. the delay time between the pump and probe. It is clear that the beating frequency is given by the energy difference between the two excitons energies $\omega_2-\omega_1$. The expression for the factors $\kappa_i$ implies that the pump energy $\Omega$ being closer to one exciton or the other can determine the relative weight of the contribution of exciton levels and thus provides tunability of the strength of coherent quantum excitonic beats. We note that the energy/frequency $\omega$ in the the ARPES spectra is the difference in energy of the material before and after photoemission i.e. $\omega = \omega_e + W - \omega_0 $ i.e. the energy that remains in the system.
\section{Application to a simple example}
\label{Application to a simple example}
To determine the signature of coherent exciton state, we assume that the coupling of excitons to the pump field is the same for both exciton levels. 
Considering the `1s' and `2s' wavefunctions with close by energies
\begin{equation}
\begin{split}
&\phi_{1s}(\bmr) = \dfrac{1}{\sqrt{\pi}a_B^{3/2}} e^{-r/a_B} \\
& \phi_{2s}(\bmr) = \dfrac{1}{4\sqrt{2\pi}a_B^{3/2}} \left( 2-\dfrac{r}{a_B}\right) e^{-r/2a_B}
\end{split}
\end{equation}
where $a_B$ is the exciton Bohr radius and their corresponding energies in terms of excitonic Rydberg
\begin{equation}
\omega_{1s} =E_g- \text{Ry} \qquad \omega_{2s} = E_g-\dfrac{\text{Ry}}{4}
\end{equation}
In Fourier space, the wavefunctions are
\begin{equation}
\begin{split}
&\phi_{1s,\bmk} = \dfrac{8\sqrt{\pi}a_B^{3/2}}{(1+k^2 a_B^2)^2} \\
& \phi_{2s,\bmk} = -32\sqrt{2\pi}a_B^{3/2} \dfrac{1-4k^2 a_B^2}{(1+4k^2 a_B^2)^2}
\end{split}
\end{equation}
Choosing the parameters typical for transition metal dichalcogenides - Energy gap $E_g$ = 0.8 eV, exciton Rydberg Ry = 100 meV, pump energy $\Omega$ = 0.73 eV, electron mass $m_e$ = 0.4794 $m_0$, hole mass $m_h$ = 0.8184 $m_0$, exciton Bohr radius $a_B$ = 11.2 $\mathrm{\AA}$, and the pump and probe temporal width $\sigma_p = \sigma = $ 30 fs. It is clear from Eq.~\ref{FullARPES} that the excitonic beat frequency is set by the difference between the exciton energy levels $\omega_{\mathrm{beat}}=\omega_1 -\omega_2$. Thus the time period of the oscillation is $T=2\pi/\omega_{\mathrm{beat}}$.\\
%
%
\begin{figure}[hbtp]
\centering
\includegraphics[width=0.45\textwidth]{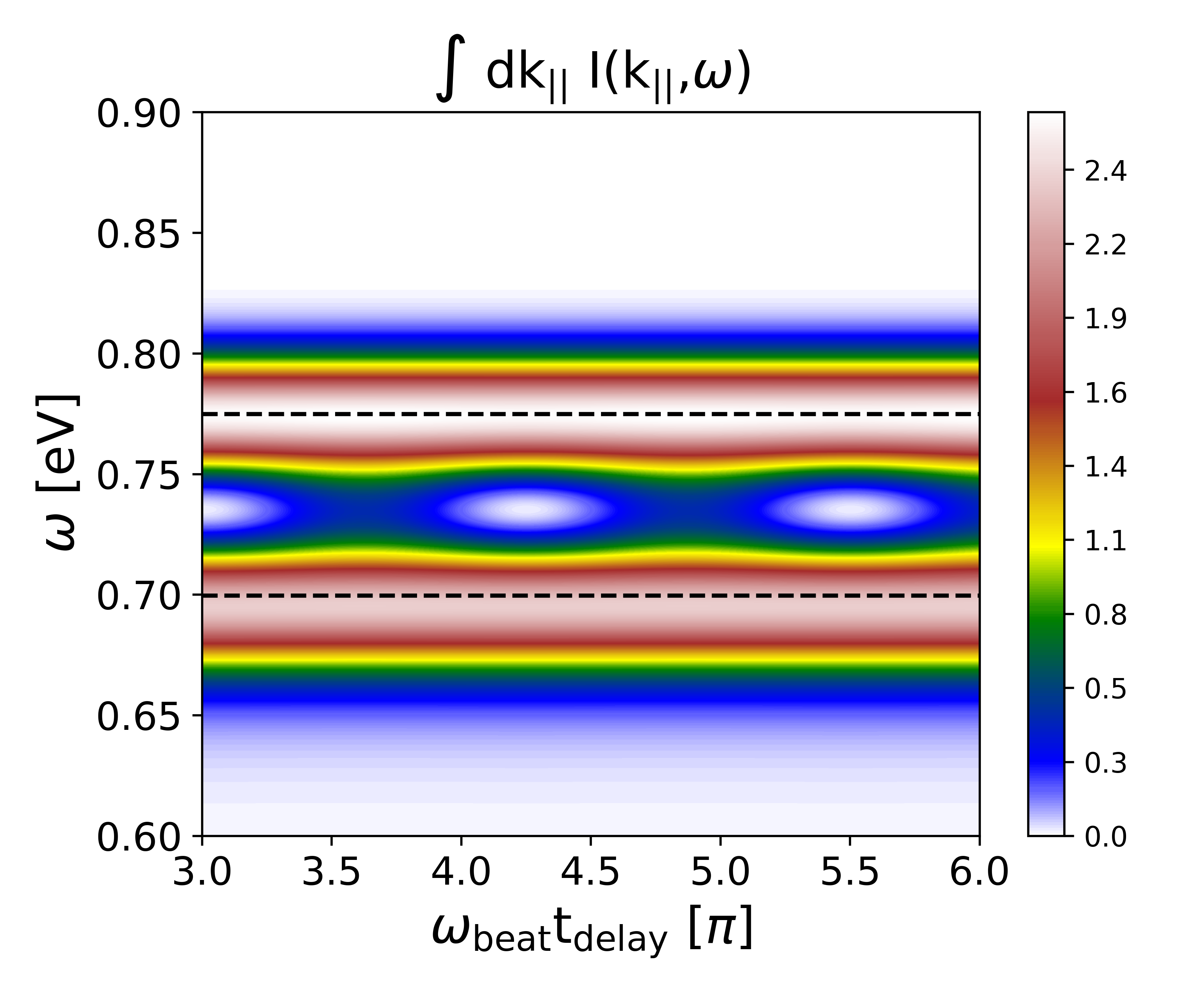}
\caption{\label{kintARPES_time}False color plot of the momentum-integrated spectra. The horizontal dashed lines are at the exciton energy levels considered. The momentum-integrated spectra clearly shows coherent excitonic quantum beats in the energy range between the exciton levels.}
\end{figure}
%
%
We note that since the larger quantum number wavefunction are more spread in real space, thus keeping the normalization of wavefunction in mind implies that the larger quantum number wavefunction are narrower in momentum and consequently have larger amplitude. We can therefore use the laser pump energy to be closer to the lower exciton level such that the contribution from the second exciton is suppressed as seen by the expression for $\kappa_i$
\begin{equation}
\kappa_{i=\{1,2\}} \approx  \sqrt{\dfrac{\pi V}{2}} E_0 G_i  \sigma_p   e^{-\sigma_p^2 (\omega_i -\Omega)^2/2} 
\end{equation}

With these two exciton levels, we apply the formalism developed in Sec.~\ref{Photoemission Theory}. We have consistent signatures of excitons to ARPES as highlighted recently\citep{rustagi2018ARPES} as seen in Fig.~\ref{ARPES}. However due the coherent excitation of the two exciton levels, there is additional interesting oscillations seen in the photoemission intensity (see Fig.~\ref{ARPES}). The most pronounced effect of the oscillations is in the energy range between the two exciton energies. This can be seen more explicitly in the momentum-integrated ARPES spectra as shown in Fig.~\ref{kint_ARPES} and Fig.~\ref{kintARPES_time}.
\section{Conclusions}
\label{Conclusions}
In this work, we have considered the signatures of coherent exciton states in photoemission measurements. The ultrafast nature of the pump pulse allows for simultaneous excitation of energetically near exciton levels, the interference of whose polarizations show up as beats. The exciton beat shows up as oscillations in photoemission intensity and this signature is most pronounced in the energy range between the two exciton energies. With coherent control of these oscillations in mind, the tunability of the pump laser pulse energy allows for dominantly exciting one exciton over the other.

\section{Acknowledgements}
The authors acknowledge the fruitful discussions with P. Kirchmann.

\bibliography{refs}
\end{document}